\documentclass[preprintnumbers,amsmath,amssymb,floatfix,12pt,prd,superscriptaddress,nofootinbib]{revtex4}
\usepackage{graphicx}
\usepackage{epsfig}
\usepackage{bm}
\usepackage{amsfonts}
\usepackage{subfigure}

\def\bea{\begin{eqnarray}}
\def\eea{\end{eqnarray}}

\begin{document}
\begin{center}
\LARGE { \bf Warm Vector Inflation
  }
\end{center}
\begin{center}
{\bf M. R. Setare\footnote{rezakord@ipm.ir} \\  V. Kamali\footnote{vkamali1362@gmail.com}}\\
 {Department of Science, University of Kurdistan,\\
Sanandaj, IRAN. }
 \\
 \end{center}
\vskip 3cm

\begin{center}
{\bf{Abstract}}\\
In this paper we introduce the "warm vector inflation" scenario. In warm inflation scenario radiation is produced during the inflation epoch and reheating is avoided. Slow-roll and perturbation parameters of this model are presented. We develop our model using intermediate inflation model. In this case, the model is compatible with observational data. We also study the model using another exact  cosmological solution, named logamediate scenario. We present slow-roll and Hubble parameters, power spectrum and tensor-scalar ratio in terms of inflaton. The model is compatible with WMAP7 and Planck observational data.

\end{center}

\newpage

\section{Introduction}
Evidence from the cosmic microwave background (CMB)
radiation indicates the early universe has an accelerating
phase i.e, the inflationary epoch. Moreover inflation theory
is one of the most compelling solution to many long-standing
problems of the standard hot big bang model, for example, the horizon,  the
flatness, and the monopole problems, among others
\cite{1',2'}. Main successful models of inflation are presented based on
weakly interacting scalar fields. Slow-roll (approximation) is
an important characteristic behavior of inflation process, but
higher spin fields induce an anisotropy in spatial segment of space and the
effective masses of these fields are usually of the order of the
Hubble scale, therefore the slow-roll inflation does not occur
\cite{3',4',5'}. Vector inflation model using an
orthogonal triple vector set and nonminimal coupling to gravity
 has been presented in Ref.\cite{1}. This model is very successful, non-minimally coupled vector fields appear to behave in precisely
the same way as a massive minimally coupled scalar field in a
flat universe.
\\
One main problem of the
inflation theory is how to attach the universe to the end of the
inflation epoch. An interesting solution of this problem is the
study of inflation in the context of warm inflation scenario \cite{n1}. In
warm model of inflation, radiation is produced during inflation period where its
energy density is kept nearly constant. This is
fulfilled by introducing the dissipation coefficient $\Gamma$.
Dissipation term may be derived, using quantum field theory (QFT) methods in a two-stage  mechanism for field interaction \cite{4nn,arj}.
 The study
of warm inflation model as a mechanism that gives an end for
vector inflation theory is motivated us to consider the warm vector inflation model.\\
In warm inflation there has to be continuously particle (photon) production.
For this to be possible, then the microscopic processes that produce
these particles must occur at a timescale ($\tau$) much faster than Hubble
expansion.  Thus the decay rates $\Gamma_i$ (not to be confused with the
dissipative coefficient) must be bigger than $H$. Also these produced
particles must thermalize. Thus the scattering processes amongst these
produced particles must occur at a rate bigger than $H$.  These
adiabatic conditions were outlined since the early warm inflation
papers, such as Ref. \cite{1-ne}. There has been considerable explicit calculations from QFT that
explicitly compute all these relevant scattering  and decay rates
in warm inflation scenario \cite{4nn,arj}.\\
In all inflationary models the scale factor is exponentially or power-law.
Exact solutions exist in both cases \cite{1',6'}.
However exact solutions may also be presented for 'intermediate'
inflationary universes in which the scale factor is given by
\cite{7',4}.
\begin{eqnarray}\label{}
a(t)=\exp(A t^f),
\end{eqnarray}
where $A$ and $f$ are two constants, $0<f<1$, and $A>0$. The
expansion of this universe is slower than standard de Sitter
inflation, but faster than other exact solutions (power-law) \cite{y}. Intermediate
inflation arises as the slow-roll solution to potential which falls
off asymptotically as a power-law in field, and may be modelled by
an exact cosmological solution.\\
In one section of the present work, we will consider warm vector inflation model in the context of "intermediate inflation". The study of this form of scale factor is motivated by string/M theory \cite{4-m}.  If we add the higher order curvature correction, which is proportional to Gauss-Bonnent (GB) term, to Einstein-Hilbert action then  a free-ghost action \cite{5-m} may be obtained. The Gauss-Bonnent interaction is leading order of the "$\alpha$" expansion to low-energy string effective action \cite{5-m} ($\alpha$ is inverse string tension). This theory may be applied for black hole solutions \cite{6-m}, acceleration of the late time universe \cite{7-m} and initial singularity problems \cite{8-m}. The GB interaction in $4d$ with dynamical dilatonic scalar coupling leads to an intermediate form of scale factor \cite{4-m,10-m,11-m}.\\
On the other hand, we would like to study our model in the context of "logamediate inflation" with scale factor
\begin{eqnarray}\label{}
a(t)=a_0 \exp(A[\ln t]^{\lambda})
\end{eqnarray}
where $\lambda >1, A>0$ \cite{12-m}. This model is converted to power-law inflation for $\lambda=1$ case. This scenario is applied for a number of scalar-tensor theories \cite{13-m}. The study of logamediate scenario is motivated by imposing weak general conditions on the cosmological models which have indefinite expansion \cite{12-m}. The effective potential of the logamediate model has been considered in dark energy models \cite{14-m}. These form of potentials are also used in supergravity, Kaluza-Klein theories and super-string models \cite{13-m,15-m}. For logamediate models the power spectrum could be either blue or red tilted \cite{16-m}. In Ref.\cite{12-m},  eight possible asymptotic scale factor solutions for cosmological dynamics would be presented. Three of these solutions are non-inflationary scale factor, another three one's of solutions give de sitter, power-low  and intermediate scale factors. Finally, two cases of these solutions have asymptotic expansion with logamediate scale factor. We will study warm vector inflation model in the context of intermediate and logamediate scenarios.
\section{The model}
One inflationary model of non-minimally coupled vector fields in an
isotropic and homogeneous universe (FRW) was presented in \cite{1}. Vector
fields in this scenario behave in the same way as a minimally coupled
scalar field. The action of this model is given by
\begin{eqnarray}\label{1}
 S=-\frac{1}{2}\int d^4x\sqrt{-g}(R+\frac{1}{2}F_{\mu\nu}F^{\mu\nu}-\frac{R}{6}A^{\mu}A_{\mu}-V(A^{\mu}A_{\mu})),
\end{eqnarray}
where
$F_{\mu\nu}=\nabla_{\mu}A_{\nu}-\nabla_{\nu}A_{\mu}=\partial_{\mu}A_{\nu}-\partial_{\nu}A_{\mu}$
and $V(A^{\mu}A_{\mu})=m^2 A^{\mu}A_{\mu}+$..., here we work in natural
unit where  $8\pi G=\hbar=c=1$. By variation of this action with
respect to the $A_{\mu}$  following field equations for
the vector fields is obtained
\begin{eqnarray}\label{2}
 \frac{1}{\sqrt{-g}}\frac{\partial}{\partial x^{\mu}}(\sqrt{-g}F^{\mu\nu})+\frac{R}{6}A^{\nu}+\frac{\partial V}{\partial A_{\nu}}=0.
\end{eqnarray}
In the FRW universe with the metric
\begin{eqnarray}\label{3}
 ds^2=dt^2-a^2(t)dx^idx_i,
\end{eqnarray}
the equation (\ref{2}), for homogeneous vector fields ($\partial_i
A_{\alpha}=0$) converts to
\begin{eqnarray}\label{4}
 \ddot{B}_i+3H\dot{B}_{i}+V'(B_j B^j)B_i=0,~~~~~~~~~~~~~~~~~~A_0=0,
\end{eqnarray}
where $B_i\equiv \frac{A_i}{a},$ $H\equiv \frac{\dot{a}}{a}$  and
prime is derivative with respect to  $B_j B^j$. Where
$V(A^{\mu}A_{\mu})=m^2 A^{\mu}A_{\mu},$ the above equation is
simplified
\begin{eqnarray}\label{5}
\ddot{B}_i+3H\dot{B}_i+m^2B_i=0,
\end{eqnarray}
which is similar to the equation for the massive minimally
coupled scalar field. Energy-momentum tensor of the
vector field model may be presented by variation of action (\ref{1}) with respect to the
metric. The components of energy-momentum tensor in term of $B_i$
have the following forms \cite{1}
\begin{eqnarray}\label{6}
T^0_0=\frac{1}{2}(\dot{B}^2_k+V(B^i B_i)),
\end{eqnarray}
\begin{eqnarray}\label{7}
T^{i}_{j}=[-\frac{5}{6}\dot{B}^2_{k}-\frac{1}{2}V(B^2)-\frac{2}{3}H\dot{B}_kB_k-\frac{1}{3}(\dot{H}+3H^2-V'(B^2))B_k^2]\delta^i_j\\
\nonumber
+\dot{B}_i\dot{B}_j+H(\dot{B}_iB_j+\dot{B}_jB_i)+(\dot{H}+3H^2-V'(B^2))B_iB_j,
  \end{eqnarray}
where the summation over index $k$ has been used. The symmetries of the FRW universe with
metric (\ref{3}) do not allow the energy-momentum tensor contains
off-diagonal components. Therefore to remove the off-diagonal
sector of energy-momentum tensor, we have to present  several
fields simultaneously. A triplet of mutually
orthogonal vector fields $B_i^{a}$ are introduced in \cite{2}. This vector fields are constrained by
orthogonality
\begin{eqnarray}\label{8}
\sum_i B_i^{a}B_i^{b} =|B|^2 \delta_a^b,
\end{eqnarray}
and
\begin{eqnarray}\label{}
\sum_a B_i^{a}B_j^{a} =|B|^2 \delta_i^j.
\end{eqnarray}
Using above relations and Eqs. (\ref{6}),(\ref{7}), the total
energy-momentum tensors of the triplet of mutually
orthogonal vector fields are given by
\begin{eqnarray}\label{10}
T^0_0=\rho =\frac{3}{2} (\dot{B}_k^2+V(|B|^2)),~~~~~~~~~~\\
\nonumber
T^i_j=-P\delta^i_j=-\frac{3}{2}(\dot{B}_k^2-V(|B|^2))\delta^i_j,
\end{eqnarray}
where $B_k$ satisfy the equation of motion (\ref{4}). In the above equation we have used an ansatz
\begin{equation}\label{11}
B_i^{(a)}=|B|\delta_i^{a},
\end{equation}
\\
In slow-roll regime when $\dot{B}_k^2\ll V(B^2)$ we have
$p\approx-\rho,$ therefore  the universe undergoes the stage of inflation.
So, inflation may be driven by non-minimally coupled vector fields
in an isotropic and homogeneous universe which behaves similar to
massive minimally coupled scalar field in FRW space-time \cite{3}.
The dynamic of  warm vector inflation in spatially flat FRW universe  is presented by these equations
\begin{eqnarray}\label{12}
\dot{\rho}+3H(P+\rho)=-\Gamma\dot{B}^2~~~~~~~~~~~~~~~~~~~~~~~~~~~~~~~~~~~~~~\\
\nonumber
\dot{\rho}_{\gamma}+4H\rho_{\gamma}=\Gamma\dot{B}^2~~~~~~~~~~~~~~~~~~~~~~~~~~~~~~~~~~~~~~~~~~~~~~\\
\nonumber
H^2=\frac{1}{3}(\frac{3}{2}(\dot{B}^2+V(B))+\rho_{\gamma})=\frac{1}{2}(\dot{B}^2+V(B))+\frac{1}{3}\rho_{\gamma}
\end{eqnarray}
where $\rho_{\gamma}$ is energy density of the radiation and
$\Gamma$ is the dissipative coefficient. In the above equations dot "." means derivative with
respect to cosmic time. Warm inflation is an example of a phase transition with dissipative effect \cite{n1,4nn}. In supercool inflation model \cite{1'}, this effect becomes important after the end of inflation (in reheating epoch,) but in warm inflation model the interactions of inflaton with other fields are important during the inflationary period. Because of these interactions the dynamic equations of inflation (\ref{12}) are completely changed. Dissipation coefficient in the above equations could be calculated, by using QFT methods in a two-stage  mechanism for field interaction \cite{4nn,arj} (for full consideration see appendix). In analogy to the QFT calculations in Refs.\cite{4nn,arj}, one might expect a form such as
\begin{eqnarray}\label{v}
\Gamma=\Gamma_0\frac{T^3}{B^2}
\end{eqnarray}
if the interaction structure is as given in the appendix. The above form is  presented as a possible type of dissipative coefficient
but there might be other forms. In the above equation $T,$ is  the temperature of thermal bath. In Sections (III) and (IV) we will use this form of dissipation coefficient. During inflation epoch the energy density  $\rho$ of inflaton
is the order of potential energy density $V(A^2)$ ($\rho\sim V$) and
dominates over the radiation energy density $\rho>\rho_{\gamma}$.
Using slow-roll approximation where
$\ddot{B}\ll(3H+\frac{\Gamma}{3})\dot{B}$ \cite{n1} and when inflation
radiation production is quasi-stable, ($\dot{\rho}_{\gamma}\ll
4H\rho_{\gamma}$, $\dot{\rho}_{\gamma}\ll\Gamma\dot{B}^2$) the
dynamic equations are reduced to
\begin{eqnarray}\label{13}
3H(1+\frac{r}{3})\dot{B}=-\frac{1}{2}V'~~~~~~~~~~~~~~~~~~\\
\nonumber
\rho_{\gamma}=\frac{3}{4}r\dot{B}^2=\frac{r}{(1+\frac{r}{3})^2}\frac{V'^2}{V}=CT^4\\
\nonumber
H^2=\frac{1}{2}V~~~~~~~~~~~~~~~~~~~~~~~~~~~~~~~~~
\end{eqnarray}
where $r=\frac{\Gamma}{3H},$ and $C=\frac{\pi^2 g^{*}}{30}$ ($g^{*}$ is the number of relativistic degree of freedom.). In the above equations prime ($'$) denotes derivative with respect to filed $B$. From above equations the temperature of thermal bath is presented
\begin{eqnarray}\label{vv}
T=[-\frac{r\dot{H}}{2C(1+\frac{r}{3})}]^{\frac{1}{4}}
\end{eqnarray}
Slow-roll parameters of the warm vector inflation are
\begin{eqnarray}\label{14}
\epsilon=-\frac{1}{H}\frac{d}{dt}\ln H=\frac{1}{12(1+\frac{r}{3})}\frac{V'^2}{V^2}~~~~~~\\
\nonumber
\eta=-\frac{\ddot{H}}{H\dot{H}}=2\epsilon-\frac{\dot{\epsilon}}{H\epsilon}~~~~~~~~~~~~~~~~~~~~\\
\nonumber
=2\epsilon+\frac{1}{3+r}\frac{V'}{V}(2\frac{V''}{V}-3\frac{V'}{V}-\frac{r'}{3+r})
\end{eqnarray}
From Eqs.(\ref{13}) and (\ref{14}), we could find a relation between
$\rho$ and $\rho_{\gamma}$.
\begin{eqnarray}\label{}
\rho_{\gamma}=\frac{3}{2}\frac{r}{3+r}\epsilon\rho
\end{eqnarray}
In high dissipative regime ($\Gamma\gg9H$) we have
\begin{eqnarray}\label{}
\rho_{\gamma}=\frac{3}{2}\epsilon\rho
\end{eqnarray}
Condition of inflation epoch ($\ddot{a}<1$) may be presented  by inequality  $\epsilon<1$.
\begin{eqnarray}\label{}
\rho>\frac{2}{3}\rho_{\gamma}
\end{eqnarray}
This is the condition for warm vector inflation era.
Warm inflation epoch ends when $\rho=\frac{2}{3}\rho_{\gamma}$. The number of e-folds
has the following form:
\begin{eqnarray}\label{}
N=\int Hdt=-\int_{\phi_{*}}^{\phi_f}\frac{(3+r)V}{V'} dt
\end{eqnarray}
where $f$ denotes the end of inflation and the epoch when the cosmological
scale exits the horizon is denoted by subscript $*$. \\
Perturbations of our model at the smallest level in spatially flat FRW background, will be studied. We will present the perturbation theory in isotropic universe using variation of inflaton $B$. In warm inflation scenario the variation of inflaton is presented by thermal fluctuation. In non-warm inflation scenarios, fluctuation of inflaton may be deriven by quantum fluctuation \cite{n2,n3}
\begin{eqnarray}\label{}
\langle\delta B\rangle_{quantum}=\frac{H^2}{2\pi}
\end{eqnarray}
In warm inflation model the thermal fluctuation provides
\begin{eqnarray}\label{15}
\langle\delta B\rangle_{thermal}=(\frac{\Gamma H T^2}{(4\pi)^3})^{\frac{1}{4}}
\end{eqnarray}
where $T,$ is the temperature of thermal bath.
 We will study  Tensor and scalar perturbations emerge during inflation epoch for warm vector inflation model. These perturbations may leave an imprint in the CMB anisotropy  and on the LSS \cite{n2,n3}. Power spectrum and a spectral index, are characteristics of each fluctuation: $\Delta_R^2(k)$ and $n_R$ for scalar perturbation, $\Delta_T^2(k)$ and $n_T$ for tensor perturbation. In warm and cool inflation models, scalar power spectrum is given by
\begin{eqnarray}\label{16}
\Delta_R^2(k)=(\frac{H}{\dot{B}}\langle\delta B\rangle)^2
\end{eqnarray}
where $k$ is co-moving wavenumber.
At the reference wavenumber $k=k_0= 0.002 Mpc^{-1},$ the combined measurement from WMAP+BAO+SN of $\Delta_R^2(k_0)$ is reported by WMAP7 data \cite{2-i}
\begin{eqnarray}\label{}
\Delta_R^2(k_0) = (2.445 \pm 0.096) \times 10^{-9}
\end{eqnarray}
In our model the power-spectrum of scalar perturbation is presented from Eqs.(\ref{15}) and (\ref{16})
\begin{eqnarray}\label{17}
\Delta_{R}^2(k)=-(\frac{\Gamma^3 T^2}{36(4\pi)^3})^{\frac{1}{2}}\frac{H^{\frac{3}{2}}}{\dot{H}}=(\frac{16\Gamma^5 T^2}{9\sqrt{2}(4\pi)^3})^{\frac{1}{2}}\frac{V^{\frac{5}{4}}}{V'^2}
\end{eqnarray}
The largest value of density perturbation is produced when $B=B_i$ \cite{n4}. Scalar spectral index of our model is presented by
\begin{eqnarray}\label{18}
n_s-1=-\frac{d\ln \Delta_{R}^2(k)}{d\ln k}
\end{eqnarray}
In warm inflation scenario, thermal fluctuations are considered instead of quantum fluctuations that generate  scalar perturbations, therefor density fluctuation of scalar perturbation is modified while the tensor perturbation shows the same spectrum as in the usual non-warm inflation \cite{n5}
\begin{eqnarray}\label{19}
\Delta^2_T=\frac{2 H^2}{\pi^2}=\frac{V}{\pi^2}
\end{eqnarray}
Spectral index $n_T$ may be found as
\begin{eqnarray}\label{}
n_T=-2\epsilon
\end{eqnarray}
From WMAP data, we could not constraint $\Delta^2_T$ directly, but the tensor-scalar ratio
\begin{eqnarray}\label{21}
R=-(\frac{144(4\pi)^3}{\Gamma^3\pi^4T^2})^{\frac{1}{2}}\dot{H}H^{\frac{1}{2}}
\end{eqnarray}
may be constrained using the WMAP+BAO+SN measurement \cite{2-i}:
\begin{eqnarray}\label{22}
R < 0.22
\end{eqnarray}
\section{Intermediate inflation}
Intermediate inflation will be studied in this section, where the scale factor of this model has the following form:
\begin{eqnarray}\label{23}
a=a_0\exp(At^f)~~~~~~0<f<1
\end{eqnarray}
where $A$ is positive constant. The number of e-folds in this case is given by using the above equation
\begin{eqnarray}\label{24}
N=\int_{t_1}^{t} H dt=A(t^{f}-t_1^{f})
\end{eqnarray}
where $t_1$ is the begining time of inflation.
\subsection{$\Gamma=\Gamma_0\frac{T^3}{B^2}$}
 From Eqs. (\ref{v},) (\ref{13},) (\ref{vv}) and (\ref{23}) we could find the Hubble parameter and scalar field
\begin{eqnarray}\label{25}
H=fA(\frac{\ln B-\ln B_0}{\omega})^{\frac{8(f-1)}{5f+2}}\\
\nonumber
B=B_0\exp(\omega t^{\frac{5f+2}{8}})~~~~~~~
\end{eqnarray}
where $\omega=(\frac{6}{\Gamma_0}(\frac{2C}{3})^{\frac{3}{4}})^{\frac{1}{2}}(\frac{8(fA)^{\frac{5}{8}}(1-f)^{\frac{1}{8}}}{5f+2})$ and $\Gamma_0=const$. Important slow-roll parameters $\epsilon$ and $\eta$ are presented
\begin{eqnarray}\label{26}
\epsilon=\frac{1-f}{fA}(\frac{\ln B-\ln B_0}{\omega})^{\frac{-8f}{5f+2}}\\
\nonumber
\eta=\frac{2-f}{fA}(\frac{\ln B-\ln B_0}{\omega})^{\frac{-8f}{5f+2}}
\end{eqnarray}
respectively.
Energy density of radiation in this case has the following form:
\begin{eqnarray}\label{}
\rho_{\gamma}=3(1-f)fA(\frac{\ln B-\ln B_0}{\omega})^{\frac{8f-2}{5f+2}}
\end{eqnarray}
Number of e-fold between two fields $B_1$ and $B$ is presented by using Eqs.(\ref{24}) and (\ref{25})
\begin{eqnarray}\label{27}
N=A[(\frac{\ln B-\ln B_0}{\omega})^{\frac{8f}{5f+2}}-(\frac{\ln B_1-\ln B_0}{\omega})^{\frac{8f}{5f+2}}]
\end{eqnarray}
At the begining of the inflation epoch where $\epsilon=1,$  the scalar field in term of constant parameters of the model is presented
\begin{eqnarray}\label{28}
B_1=B_0\exp(\omega(\frac{1-f}{fA})^{\frac{5f+2}{8f}})
\end{eqnarray}
From  above equations we could find the inflaton $B(t)$ in term of number of e-folds
\begin{eqnarray}\label{29}
B=B_0\exp(\omega(\frac{N}{A}+\frac{1-f}{fA})^{\frac{5f+2}{8f}})
\end{eqnarray}
\begin{figure}[h]
\centering
  \includegraphics[width=10cm]{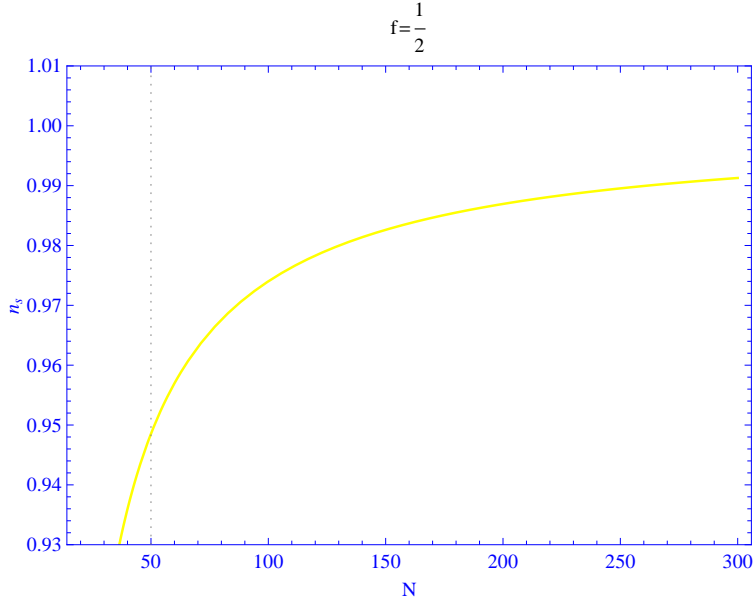}
  \caption{In this graph we plot the spectral index $n_s$  versus  the number of e-folds $N$ for intermediate scenario ($A=1$, $f=\frac{1}{2}$).}
 \label{fig:F3}
\end{figure}
Perturbation parameters versus the scalar field $B,$ and constant parameters of intermediate scenario are presented for warm vector inflation model.
In slow-roll limit the power-spectrum of scalar perturbation could be found using Eqs.(\ref{15}), (\ref{16}) and (\ref{29})
\begin{eqnarray}\label{30}
\Delta_R^2=(\frac{\Gamma_0^3}{36(4\pi)^3})^{\frac{1}{2}}(\frac{3^{11}(fA)^{15}(1-f)^3}{(2C)^{11}})^{\frac{1}{8}}B^3(\frac{\ln B-\ln B_0}{\omega})^{\frac{15f-18}{5f+2}}~~~~~~~~~~~~~~~~~~~~~~~\\
\nonumber
=(\frac{\Gamma_0^3}{36(4\pi)^3})^{\frac{1}{2}}(\frac{3^{11}(fA)^{15}(1-f)^3}{(2C)^{11}})^{\frac{1}{8}}B_0^{-3}
\exp(-3\omega(\frac{N}{A}+\frac{1-f}{fA})^{\frac{5f+2}{8f}})(\frac{N}{A}+\frac{1-f}{fA})^{\frac{15f-18}{8f}}
\end{eqnarray}
Another important perturbation parameter is spectral index $n_s$ which is given by
\begin{eqnarray}\label{31}
n_s-1\simeq\frac{15f-18}{8fA}(\frac{\ln B-\ln B_0}{\omega})^{\frac{-8f}{5f+2}}=\frac{15f-18}{8fA}(\frac{N}{A}+\frac{1-f}{fA})^{-1}
\end{eqnarray}
In Fig.(1),  the spectral index $n_s$  in term of the number of e-folds is plotted (where $f=\frac{1}{2}$).
From this graph, it is observed that the model is compatible with observational data \cite{2-i,2-m}, ($N\simeq 60$ case  leads to $0.956<n_s< 0.97$).
Tensor power spectrum and its spectral index are given by
\begin{eqnarray}\label{32}
\Delta^2_T=\frac{2(fA)^2}{\pi^2}(\frac{\ln B-\ln B_0}{\omega})^{\frac{16(f-1)}{5f+2}}=\frac{2(fA)^2}{\pi^2}(\frac{N}{A}+\frac{1-f}{fA})^{\frac{2(f-1)}{f}}\\
\nonumber
n_T=-\frac{2-2f}{fA}(\frac{\ln B-\ln B_0}{\omega})^{\frac{-8f}{5f+2}}~~~~~~~~~~~~~~~~~~~~~~~~~~~~~~~~~~~~~
\end{eqnarray}
Tensor-scalar ratio has the following form
\begin{eqnarray}\label{33}
R=(\frac{144(4\pi)^3(fA)^4}{\Gamma_0^3\pi^4})^{\frac{1}{2}}(\frac{3^{11}(fA)^{15}(1-f)^3}{(2C)^{11}})^{-\frac{1}{8}} B^3(\frac{\ln B-\ln B_0}{\omega})^{\frac{f+2}{5f+2}}~~~~~~~~~~~~~~~~~~~~~~~~~~~~~~~~~~~~~~~\\
\nonumber
=(\frac{144(4\pi)^3(fA)^4}{\Gamma_0^3\pi^4})^{\frac{1}{2}}(\frac{3^{11}(fA)^{15}(1-f)^3}{(2C)^{11}})^{-\frac{1}{8}}B_0^3\exp(-3\omega(\frac{N}{A}+\frac{1-f}{fA})^{\frac{5f+2}{8f}})(\frac{N}{A}+\frac{1-f}{fA})^{\frac{f+2}{8f}}~~~~~~\\
\nonumber
=(\frac{144(4\pi)^3(fA)^4}{\Gamma_0^3\pi^4})^{\frac{1}{2}}(\frac{3^{11}(fA)^{15}(1-f)^3}{(2C)^{11}})^{-\frac{1}{8}}B_0^3\exp(-3\omega(\frac{18-15f}{8fA(1-n_s)})^{\frac{5f+2}{8f}})(\frac{18-15f}{8fA(1-n_s)})^{\frac{f+2}{8f}}
\end{eqnarray}
\begin{figure}[h]
\centering
  \includegraphics[width=10cm]{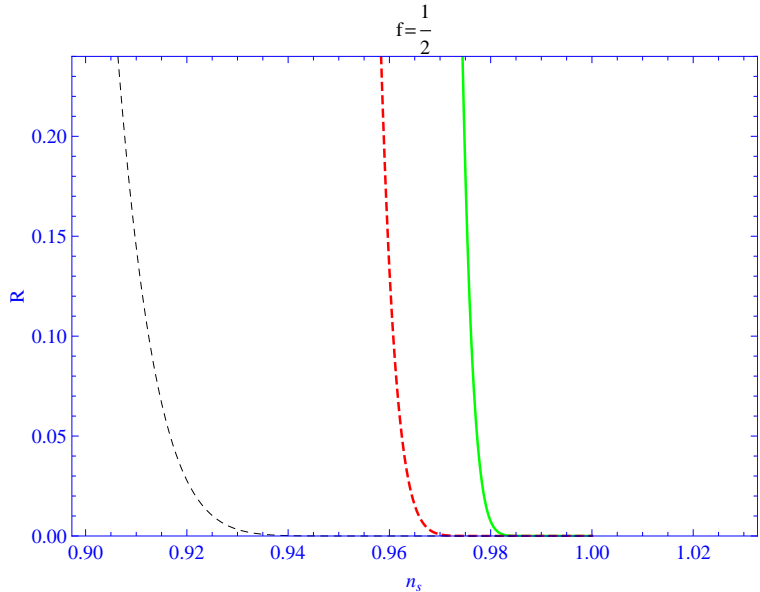}
  \caption{In this graph we plot the scalar-tensor ratio $R$   in term of the spectral index $n_s$ for intermediate scenario ( $\Gamma_0=4$ with green line, $\Gamma_0=1$ with red dashed line, and $\Gamma_0=0.25$ with black dashed line, $A=1,C=70,f=\frac{1}{2}, B_0\propto C^{-\frac{1}{12}}$).}
 \label{fig:F3}
\end{figure}
The graph of the tensor-scalar ratio $R$ in term of spectral index $n_s$  is presented in Fig.(2). Standard case $n_s\simeq 0.96$, my be found in  $0.01<R<0.22,$ for important case $\Gamma_0=1$ \cite{arj},
 which is agree with observational data \cite{2-i,2-m}.
\subsection{$\Gamma=\Gamma_1=const$}
From Eqs.(\ref{13}) and (\ref{23})  the Hubble parameter and the scalar field could be found
\begin{eqnarray}\label{251}
H=fA(\frac{B-B_0}{\omega})^{\frac{2(f-1)}{2f-1}}\\
\nonumber
B=B_0+\omega t^{\frac{2f-1}{2}}~~~~~~~
\end{eqnarray}
where $\omega=(\frac{24 (fA)^2(1-f)}{\Gamma_1(2f-1)^2})$ and $\Gamma_1=const$. Important slow-roll parameters $\epsilon$ and $\eta$ are given by
\begin{eqnarray}\label{}
\epsilon=\frac{1-f}{fA}(\frac{B-B_0}{\omega})^{\frac{2f}{1-2f}}\\
\nonumber
\eta=\frac{2-f}{fA}(\frac{B-B_0}{\omega})^{\frac{2f}{1-2f}}
\end{eqnarray}
respectively.
The energy density of radiation in this case has this form:
\begin{eqnarray}\label{}
\rho_{\gamma}=3(1-f)fA(\frac{B-B_0}{\omega})^{\frac{4-2f}{1-2f}}
\end{eqnarray}
The number of e-fold between two fields $B_1$ and $B$ is presented (We have used Eqs.(\ref{24}) and (\ref{251}))
\begin{eqnarray}\label{}
N=A[(\frac{B-B_0}{\omega})^{\frac{2f}{2f-1}}-(\frac{B_1-B_0}{\omega})^{\frac{2f}{2f-1}}]
\end{eqnarray}
At the beginning of the inflation epoch where $\epsilon=1,$  the scalar field in terms of constant parameters of the model is given by
\begin{eqnarray}\label{}
B_1=B_0+\omega(\frac{1-f}{fA})^{\frac{2f-1}{2f}}
\end{eqnarray}
Using  above equations we could find the inflaton $B(t)$ in terms of the number of e-folds
\begin{eqnarray}\label{291}
B=B_0+\omega(\frac{N}{A}+\frac{1-f}{fA})^{\frac{2f-1}{2f}}
\end{eqnarray}
\begin{figure}[h]
\centering
  \includegraphics[width=10cm]{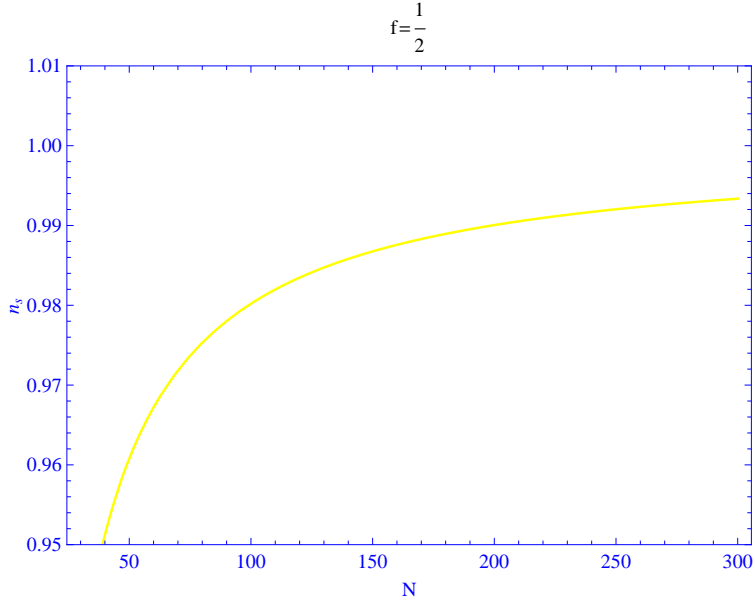}
  \caption{In this graph we plot the spectral index $n_s$  versus  the number of e-folds $N$ for intermediate scenario ($A=1$, $f=\frac{1}{2}$.}
 \label{fig:F3}
\end{figure}
The perturbation parameters  versus the scalar field $B,$  and constant parameters of intermediate scenario in this case are presented for warm vector inflation model.
In slow-roll limit, by using Eqs. (\ref{17}) and (\ref{291}), we could find the power-spectrum of scalar perturbation
\begin{eqnarray}\label{}
\Delta_R^2=(\frac{\Gamma_1^3 3^{\frac{1}{2}}}{36(2C)^{\frac{1}{2}}(4\pi)^3(fA)^{\frac{1}{2}}(1-f)^{\frac{3}{2}}})^{\frac{1}{2}}
(\frac{B-B_0}{\omega})^{\frac{3f}{2(2f-1)}}\\
\nonumber
(\frac{\Gamma_1^3 3^{\frac{1}{2}}}{36(2C)^{\frac{1}{2}}(4\pi)^3(fA)^{\frac{1}{2}}(1-f)^{\frac{3}{2}}})^{\frac{1}{2}}(\frac{N}{A}+\frac{f+1}{2f})^{\frac{3}{4}}
\end{eqnarray}
Another important perturbation parameter is spectral index $n_s$ which is presented by
\begin{eqnarray}\label{31}
n_s-1=-\frac{3}{4A}(\frac{B-B_0}{\omega})^{\frac{2f}{1-2f}}=-\frac{3}{4A}(\frac{N}{A}+\frac{1-f}{fA})^{-1}
\end{eqnarray}
In Fig.(3),  we plot the spectral index $n_s$  in term of the number of e-folds  (where $f=\frac{1}{2}$).
From this graph, it is observed that the model is compatible with observational data \cite{2-i,2-m}, ($N\simeq 50$ case  leads to $n_s\simeq 0.96$).
The tensor power spectrum and its spectral index are presented by
\begin{eqnarray}\label{32}
\Delta^2_T=\frac{2(fA)^2}{\pi^2}(\frac{B-B_0}{\omega})^{\frac{4(f-1)}{2f-1}}=\frac{2(fA)^2}{\pi^2}(\frac{N}{A}+\frac{1-f}{fA})^{\frac{2(f-1)}{f}}\\
\nonumber
n_T=-\frac{2-2f}{fA}(\frac{B-B_0}{\omega})^{\frac{2f}{1-2f}}~~~~~~~~~~~~~~~~~~~~~~~~~~~~~~~~~~~~~
\end{eqnarray}
The tensor-scalar ratio has the following form:
\begin{eqnarray}\label{33}
R=(\frac{144(2C)^{\frac{1}{2}}(4\pi)^3(fA)^{\frac{5}{2}}(1-f)^{\frac{3}{2}}}{\Gamma_1^3 \pi^4 3^{\frac{1}{2}}})^{\frac{1}{2}}
(\frac{B-B_0}{\omega})^{\frac{5f-8}{2(2f-1)}}~~~\\
\nonumber
=(\frac{144(2C)^{\frac{1}{2}}(4\pi)^3(fA)^{\frac{5}{2}}(1-f)^{\frac{3}{2}}}{\Gamma_1^3 \pi^4 3^{\frac{1}{2}}})^{\frac{1}{2}}
(\frac{N}{A}+\frac{1-f}{fA})^{\frac{5f-8}{4f}}\\
\nonumber
=(\frac{144(2C)^{\frac{1}{2}}(4\pi)^3(fA)^{\frac{5}{2}}(1-f)^{\frac{3}{2}}}{\Gamma_1^3 \pi^4 3^{\frac{1}{2}}})^{\frac{1}{2}}
(\frac{4A}{3}(1-n_s))^{\frac{8-5f}{4f}}
\end{eqnarray}
\begin{figure}[h]
\centering
  \includegraphics[width=10cm]{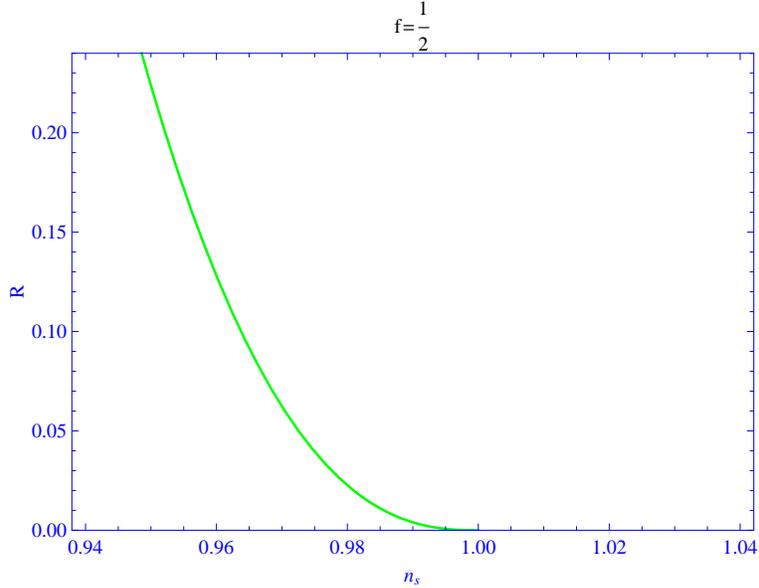}
  \caption{In this graph we plot the scalar-tensor ratio $R$   in term of the spectral index $n_s$ for intermediate scenario ($A=1,f=\frac{1}{2}$, $\Gamma_1\propto C^{\frac{1}{6}}$).}
 \label{fig:F3}
\end{figure}
The graph of the tensor-scalar ratio $R$ in terms of spectral index $n_s$  is given by Fig.(4). The standard case $n_s\simeq 0.96$, my be found in  $0.01<R<0.22,$
 which is agree with WMAP data \cite{2-i,2-m}.
\section{Logamediate inflation}
In this section we study warm vector field logamediate inflation where the scale factor has this form
\begin{eqnarray}\label{34}
a(t)=a_0\exp(A[\ln t]^{\lambda}) ~~~~~~\lambda>1,
\end{eqnarray}
$A,$ is a constant parameter. The number of e-folds may be found, using the above equation
\begin{eqnarray}\label{35}
N=\int_{t1}^{t} H dt=A[(\ln t)^{\lambda}-(\ln t_1)^{\lambda}]
\end{eqnarray}
where $t_1$ denotes the begining time of inflation epoch.
\subsection{$\Gamma=\Gamma_0\frac{T^3}{B^2}$}
 Using Eqs.(\ref{v},) (\ref{13},) (\ref{vv},) and (\ref{34}) we may find the inflaton $B$
\begin{eqnarray}\label{36}
\ln B-\ln B_0=\overline{\omega}\Xi(t)
\end{eqnarray}
where $\overline{\omega}=(\frac{6}{\Gamma_0}(\frac{2C}{3})^{\frac{3}{4}})^{\frac{1}{2}}((-4)^{5\lambda +3}(\lambda A)^5)^{\frac{1}{8}},$ and $\Xi=\gamma[\frac{5\lambda+3}{8},\frac{\ln t}{4}]$ ($\gamma[a,t]$ is incomplete gamma function \cite{gamma}). Potential  in term of scalar field $B$ is presented as:
\begin{eqnarray}\label{37}
V=\frac{2\lambda^2 A^2[\ln(\Xi^{-1}(\frac{\ln B-\ln B_0}{\overline{\omega}}))]^{2\lambda-2}}{(\Xi^{-1}(\frac{\ln B-\ln B_0}{\overline{\omega}}))^2}
\end{eqnarray}

Slow-roll parameters of the model in this case are given by
\begin{eqnarray}\label{38}
\epsilon=\frac{[\ln \Xi^{-1}(\frac{\ln B-\ln B_0}{\overline{\omega}})]^{1-\lambda}}{\lambda A}\\
\nonumber
\eta=\frac{2[\ln \Xi^{-1}(\frac{\ln B-\ln B_0}{\overline{\omega}})]^{1-\lambda}}{\lambda A}
\end{eqnarray}
The number of e-folds between two fields $B_1$ and $B(t)$ may be determined, using Eqs.(\ref{35}) and (\ref{36}).
\begin{eqnarray}\label{39}
N=A((\ln \Xi^{-1}(\frac{\ln B-\ln B_0}{\overline{\omega}}))^{\lambda}-(\ln \Xi^{-1}(\frac{\ln B_1-\ln B_0}{\overline{\omega}}))^{\lambda})\\
\nonumber
=A((\ln \Xi^{-1}(\frac{\ln B-\ln B_0}{\overline{\omega}}))^{\lambda}-(\lambda A)^{\frac{\lambda}{1-\lambda}})
\end{eqnarray}
where $B_1$ is inflaton at the begining of inflation epoch when ($\epsilon=1$). Inflaton field in the inflation period could be obtained in term of number of e-folds by using above equation
\begin{eqnarray}\label{40}
B=B_0\exp(\overline{\omega}\Xi[\exp([\frac{N}{A}+(\lambda A)^{\frac{\lambda}{1-\lambda}}]^{\frac{1}{\lambda}})])
\end{eqnarray}
Power spectrum and tensor-scalar ratio in this case are presented
\begin{eqnarray}\label{41}
\Delta_R^2=(\frac{\Gamma_0^3}{36(4\pi)^3})^{\frac{1}{2}}(\frac{3^{11}(\lambda A)^{15}}{(2C)^{11}})^{\frac{1}{8}}B_0^{-3}\exp(-\frac{15}{8}[\frac{N}{A}+(\lambda A)^{\frac{\lambda}{1-\lambda}}]^{\frac{1}{\lambda}})\\
\nonumber
(\frac{N}{A}+(\lambda A)^{\frac{\lambda}{1-\lambda}})^{15\frac{\lambda-1}{8\lambda}}
\exp(-3\overline{\omega}\Xi(\exp([\frac{N}{A}+(\lambda A)^{\frac{\lambda}{1-\lambda}}]^{\frac{1}{\lambda}})))\\
\nonumber
\Delta_T^2=\frac{2\lambda^2 A^2}{\pi^2}\frac{(\ln \Xi^{-1}(\frac{\ln B-\ln B_0}{\overline{\omega}}))^{2-2\lambda}}{(\Xi^{-1}(\frac{\ln B-\ln B_0}{\overline{\omega}}))^2}~~~~~~~~~~~~~~~~~~~~~~~~~~~~~\\
\nonumber
=\frac{2\lambda^2 A^2}{\pi^2}\exp[-2(\frac{N}{A}+(\lambda A)^{\frac{\lambda}{1-\lambda}})^{\frac{1}{\lambda}}](\frac{N}{A}+(\lambda A)^{\frac{\lambda}{1-\lambda}})^{\frac{2-2\lambda}{\lambda}}
\end{eqnarray}
Spectral indices for our model have the following forms
\begin{eqnarray}\label{42}
n_s-1=-\frac{15(\lambda -1)}{8\lambda A}[\frac{N}{A}+(\lambda A)^{\frac{\lambda}{1-\lambda}}]^{-1}~~~~~\\
\nonumber
n_T=-2\frac{[\ln \Xi^{-1}(\frac{\ln B-\ln B_0}{\overline{\omega}})]^{1-\lambda}}{\lambda A}~~~~~~~~~~~~~~~~
\end{eqnarray}
\begin{figure}[h]
\begin{minipage}[b]{1\textwidth}
\subfigure[\label{fig1a} ]{ \includegraphics[width=.37\textwidth]%
{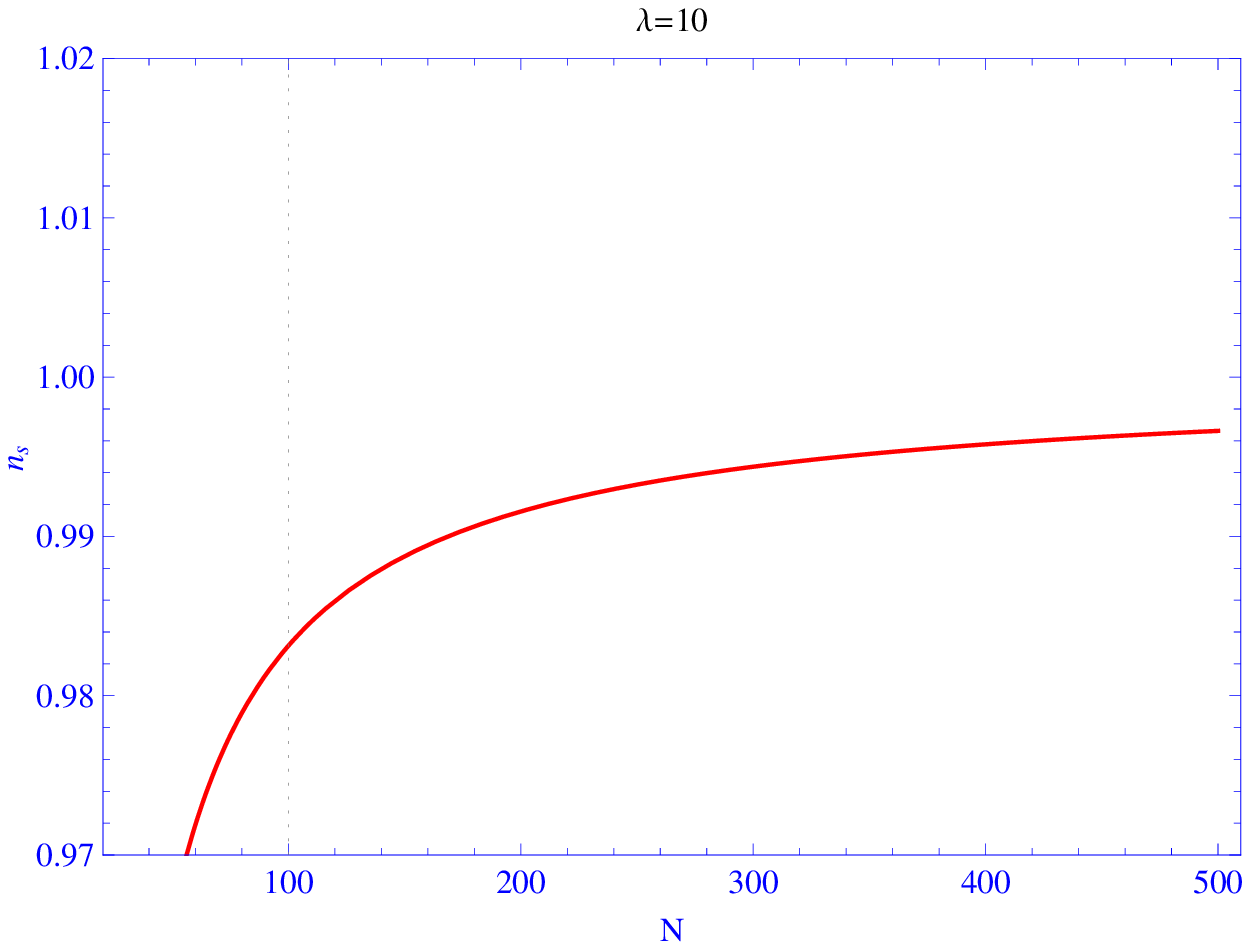}} \hspace{.2cm}
\subfigure[\label{fig1b} ]{ \includegraphics[width=.37\textwidth]%
{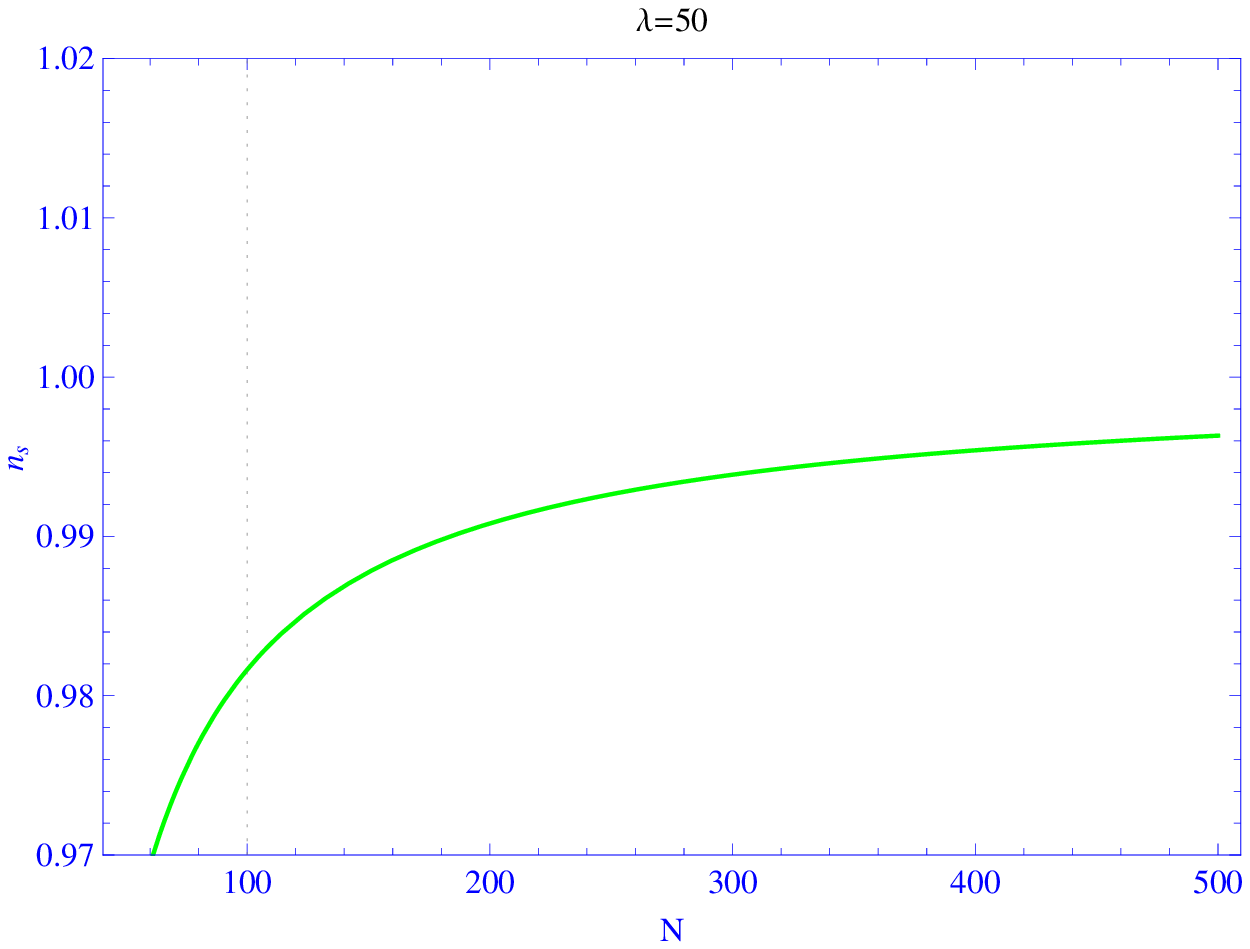}}
\end{minipage}
\caption{ Spectral index $n_s$ in term of number of e-folds $N$: (a) for $\lambda=10$ and (b) for $\lambda=50$.}
\end{figure}
In Fig.(5), the spectral index $n_s$ in term of number of e-folds is plotted  (for $\lambda=10$, $\lambda=50$,  cases). We may see the small values of number of e-folds are assured for large values of $\lambda$ parameter.
We could find the tensor-scalar ratio in term of number of e-folds and spectral index $n_s$
\begin{eqnarray}\label{43}
R=(\frac{144(4\pi)^3}{\Gamma_0^3\pi^4 })^{\frac{1}{2}}(\frac{(2C)^{11}(\lambda A)}{3^{11}})^{\frac{1}{8}}B_0^3\exp[-\frac{1}{8}(\frac{N}{A}+(\lambda A)^{\frac{\lambda}{1-\lambda}})^{\frac{1}{\lambda}}]\\
\nonumber
(\frac{N}{A}+(\lambda A)^{\frac{\lambda}{1-\lambda}})^{-\frac{31(\lambda-1)}{8\lambda}}\exp(3\overline{\omega}\Xi(\exp([\frac{N}{A}+(\lambda A)^{\frac{\lambda}{1-\lambda}}]^{\frac{1}{\lambda}})))\\
\nonumber
=(\frac{144(4\pi)^3}{\Gamma_0^3\pi^4 })^{\frac{1}{2}}(\frac{(2C)^{11}(\lambda A)}{3^{11}})^{\frac{1}{8}}B_0^3\exp(-\frac{1}{8}(\frac{15(\lambda-1)}{ 8\lambda A(1-n_s)})^{\frac{1}{\lambda}})\\
\nonumber
(\frac{15(\lambda-1)}{8\lambda A(1-n_s)})^{-\frac{31(\lambda-1)}{8\lambda}}\exp(3\overline{\omega}\Xi[\exp((\frac{15(\lambda-1)}{8\lambda A(1-n_s)})^{\frac{1}{\lambda}})])
\end{eqnarray}
\begin{figure}[h]
\begin{minipage}[b]{1\textwidth}
\subfigure[\label{fig1a} ]{ \includegraphics[width=.37\textwidth]%
{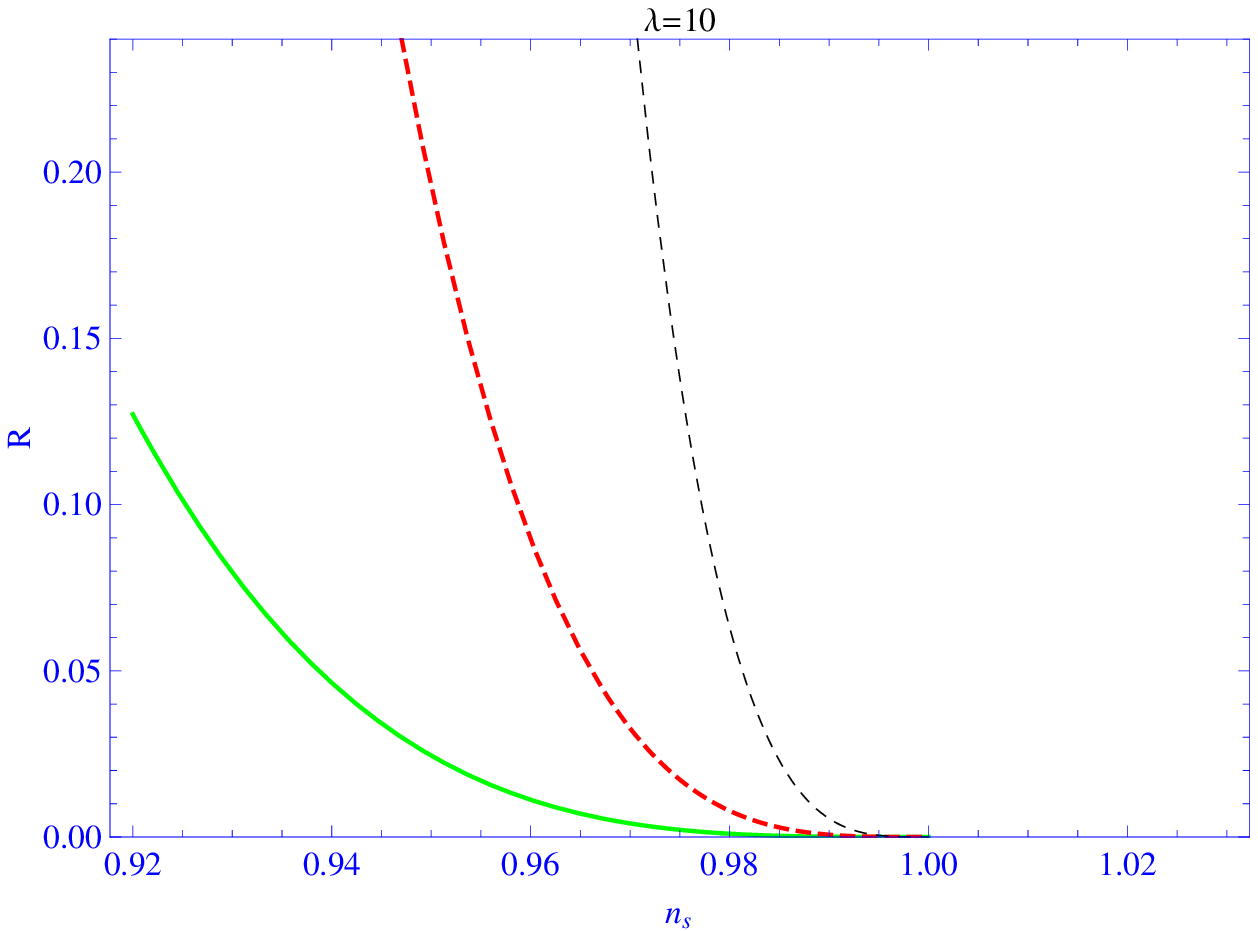}} \hspace{.2cm}
\subfigure[\label{fig1b} ]{ \includegraphics[width=.37\textwidth]%
{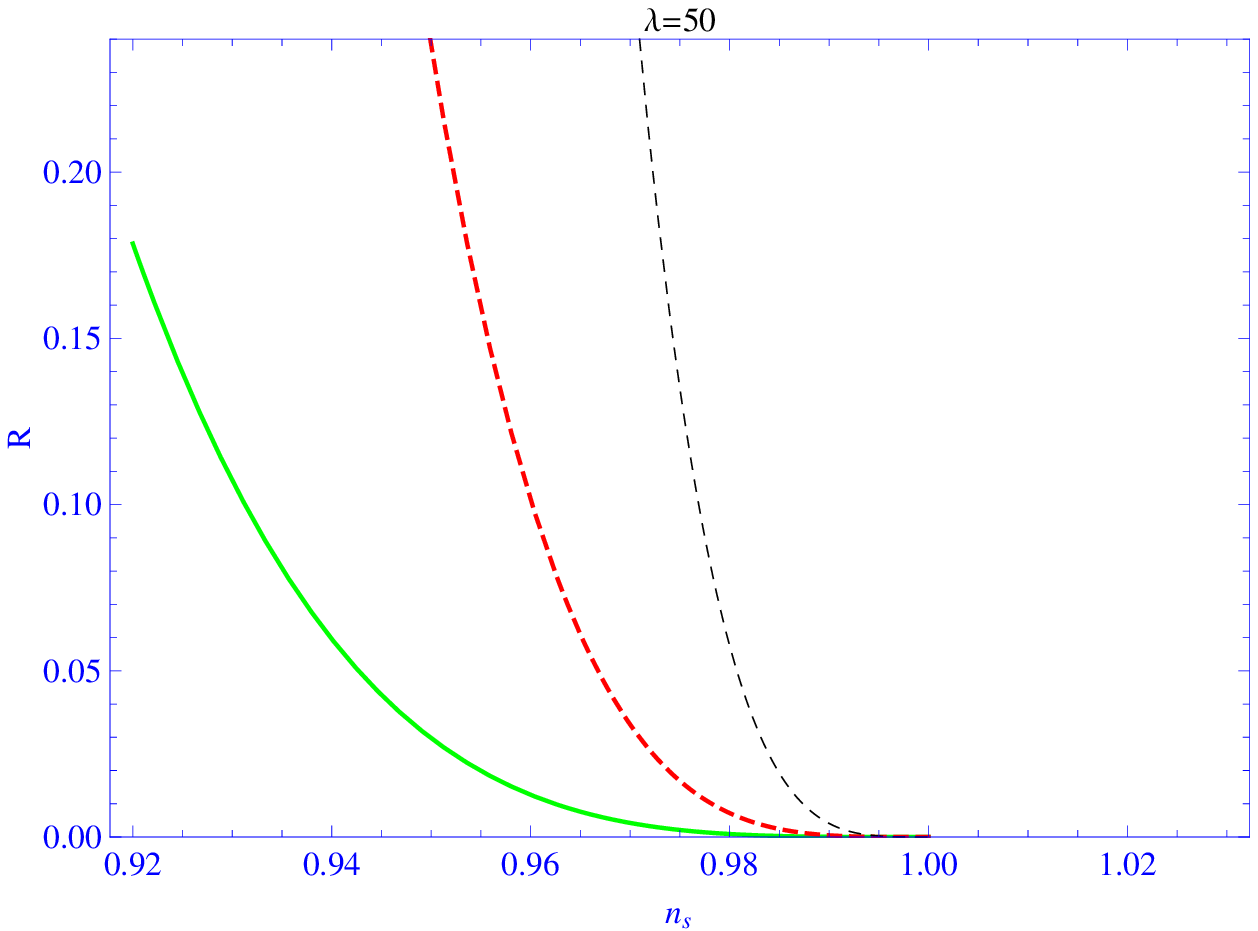}}
\end{minipage}
\caption{Scalar-tensor ratio in term of spectral index $n_s$: (a) for $\lambda=10$ and (b) for $\lambda=50$ ( $\Gamma_0=4$ with green line, $\Gamma_0=1$ with red dashed line, and $\Gamma_0=0.25$ with black dashed line, $A=1,C=70,f=\frac{1}{2}, B_0\propto C^{-\frac{1}{12}}$).}
\end{figure}
In Fig.(6), the tensor-scalar ratio  versus the spectral index is plotted  for $\lambda=10$, $\lambda=50$. We find the model is compatible with observational data (for $\Gamma_0=1$ case)\cite{2-i,2-m}.
\subsection{$\Gamma=\Gamma_1$}
Using Eqs.(\ref{13}) and (\ref{34}) we could find the inflaton field $B$
\begin{eqnarray}\label{361}
B-B_0=\overline{\omega}\Xi(t)
\end{eqnarray}
where $\overline{\omega}=(\frac{6\lambda^2 A^2}{\Gamma_1})^{\frac{1}{2}}\frac{(\lambda-1)!}{2^{\lambda-1}},$ and $\Xi=\gamma[\lambda,\frac{\ln t}{2}]$ ($\gamma[a,t]$ is incomplete gamma function \cite{gamma}). Potential  in terms of scalar field $B$ is given by
\begin{eqnarray}\label{}
V=2(\frac{\lambda A[\ln(\Xi^{-1}(\frac{B-B_0}{\overline{\omega}}))]^{\lambda-1}}{\Xi^{-1}(\frac{B-B_0}{\overline{\omega}})})^2
\end{eqnarray}
The slow-roll parameters of the model in this case are presented as
\begin{eqnarray}\label{}
\epsilon=\frac{[\ln \Xi^{-1}(\frac{B-B_0}{\overline{\omega}})]^{1-\lambda}}{\lambda A}\\
\nonumber
\eta=\frac{2[\ln \Xi^{-1}(\frac{B-B_0}{\overline{\omega}})]^{1-\lambda}}{\lambda A}
\end{eqnarray}
Using Eqs.(\ref{35}) and (\ref{361}), we could determine the number of e-folds between two fields $B_1$ and $B(t)$.
\begin{eqnarray}\label{}
N=A((\ln \Xi^{-1}(\frac{B-B_0}{\overline{\omega}}))^{\lambda}-(\ln \Xi^{-1}(\frac{B_1-B_0}{\overline{\omega}}))^{\lambda})\\
\nonumber
=A((\ln \Xi^{-1}(\frac{B-B_0}{\overline{\omega}}))^{\lambda}-(\lambda A)^{\frac{\lambda}{1-\lambda}})
\end{eqnarray}
where $B_1$ is inflaton at the begining of inflation epoch where ($\epsilon=1$). Inflaton field in the inflation period may be obtained in terms of number of e-folds by using above equation
\begin{eqnarray}\label{40}
B=B_0+\overline{\omega}\Xi[\exp([\frac{N}{A}+(\lambda A)^{\frac{\lambda}{1-\lambda}}]^{\frac{1}{\lambda}})]
\end{eqnarray}
The power spectrum and tensor-scalar ratio in this case are given by
\begin{eqnarray}\label{41}
\Delta_R^2=(\frac{(\lambda A)^{\frac{3}{2}}3^{\frac{1}{2}}\Gamma_1^3}{36(4\pi)^3(2C)^{\frac{1}{2}}})(\ln \Xi^{-1}(\frac{B-B_0}{\overline{\omega}}))^{\frac{3(\lambda-1)}{4}}~~~~~~~~~~~~~~~~~~~\\
\nonumber
=(\frac{(\lambda A)^{\frac{3}{2}}3^{\frac{1}{2}}\Gamma_1^3}{36(4\pi)^3(2C)^{\frac{1}{2}}})(\frac{N}{A}+(\lambda A)^{\frac{\lambda}{1-\lambda}})^{\frac{3(\lambda-1)}{4\lambda}}~~~~~~~~~~~~~~~~~~~~~\\
\nonumber
\Delta_T^2=\frac{2\lambda^2 A^2}{\pi^2}\frac{(\ln \Xi^{-1}(\frac{B-B_0}{\overline{\omega}}))^{2\lambda-2}}{(\Xi^{-1}(\frac{B-B_0}{\overline{\omega}}))^2}~~~~~~~~~~~~~~~~~~~~~~~~~~~~~\\
\nonumber
=\frac{2\lambda^2 A^2}{\pi^2}\exp[-2(\frac{N}{A}+(\lambda A)^{\frac{\lambda}{1-\lambda}})^{\frac{1}{\lambda}}](\frac{N}{A}+(\lambda A)^{\frac{\lambda}{1-\lambda}})^{\frac{2\lambda-2}{\lambda}}
\end{eqnarray}
The spectral indices for our model have the following forms
\begin{eqnarray}\label{42}
n_s-1=-\frac{3(\lambda-1)}{4\lambda A}[\frac{N}{A}+(\lambda A)^{\frac{\lambda}{1-\lambda}}]^{-1}~~~~~\\
\nonumber
n_T=-2\frac{(\ln\Xi^{-1}(\frac{B-B_0}{\overline{\omega}}))^{1-\lambda}}{\lambda A}~~~~~~~~~~~~~~~~
\end{eqnarray}
\begin{figure}[h]
\begin{minipage}[b]{1\textwidth}
\subfigure[\label{fig1a} ]{ \includegraphics[width=.37\textwidth]%
{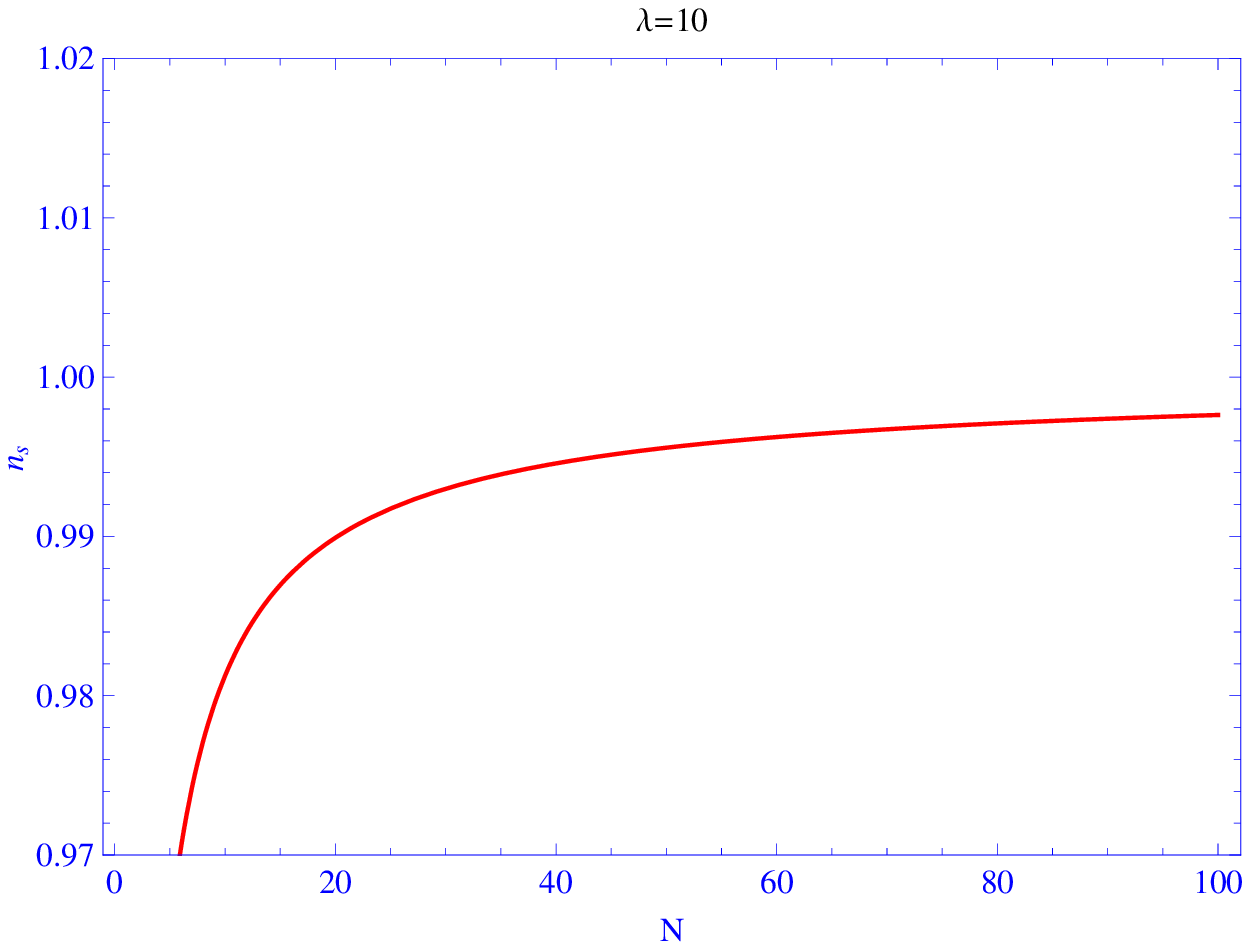}} \hspace{.2cm}
\subfigure[\label{fig1b} ]{ \includegraphics[width=.37\textwidth]%
{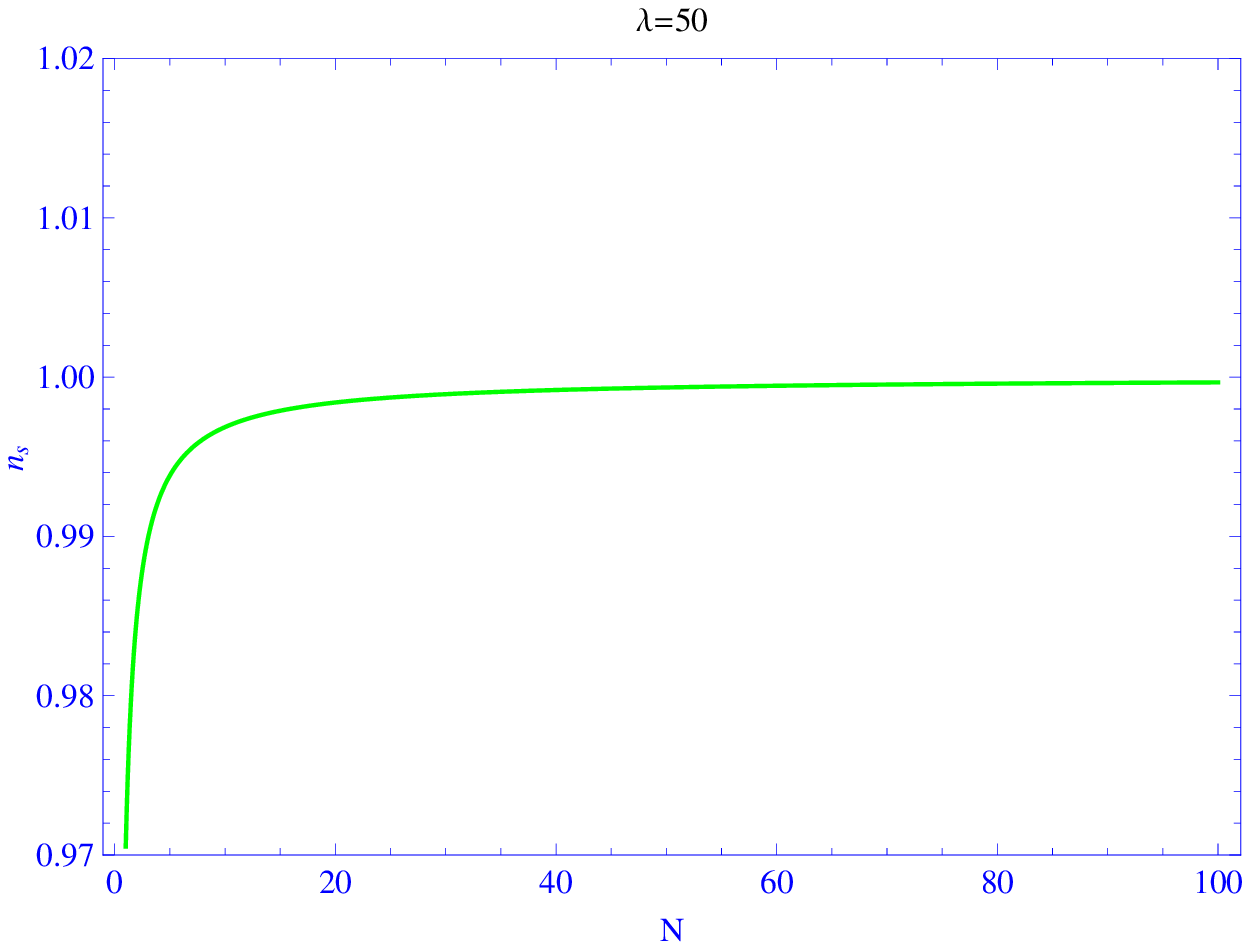}}
\end{minipage}
\caption{ Spectral index $n_s$ in term of number of e-folds $N$: (a) for $\lambda=10$ and (b) for $\lambda=50$.}
\end{figure}
In Fig.(7), the spectral index $n_s$ in terms of number of e-folds is plotted  (for $\lambda=10$, $\lambda=50$,  cases). We could see the small values of the number of e-folds are assured for large values of $\lambda$ parameter.
We also could find the tensor-scalar ratio in terms of number of e-folds and spectral index $n_s$
\begin{eqnarray}\label{43}
R=(\frac{144(2C)^{\frac{1}{2}}(4\pi)^3(\lambda A)^{\frac{5}{2}}}{\Gamma_1^3\pi^4 3^{\frac{1}{2}}})^{\frac{1}{2}}
(\Xi^{-1}(\frac{B-B_0}{\overline{\omega}}))^{-2}(\ln\Xi^{-1}(\frac{B-B_0}{\overline{\omega}}))^{\frac{5(\lambda-1)}{4}}\\
\nonumber
=(\frac{144(2C)^{\frac{1}{2}}(4\pi)^3(\lambda A)^{\frac{5}{2}}}{\Gamma_1^3\pi^4 3^{\frac{1}{2}}})^{\frac{1}{2}}
\exp[-2(\frac{N}{A}+(\lambda A)^{\frac{\lambda}{1-\lambda}})^{\frac{1}{\lambda}}](\frac{N}{A}+(\lambda A)^{\frac{\lambda}{1-\lambda}})^{\frac{5(\lambda-1)}{4\lambda}}\\
\nonumber
=(\frac{144(2C)^{\frac{1}{2}}(4\pi)^3(\lambda A)^{\frac{5}{2}}}{\Gamma_1^3\pi^4 3^{\frac{1}{2}}})^{\frac{1}{2}}
\exp(-2(\frac{4\lambda A}{3(\lambda-1)}(1-n_s))^{\frac{-1}{\lambda}})(\frac{4\lambda A}{3(\lambda-1)}(1-n_s))^{\frac{5(1-\lambda)}{4\lambda}}
\end{eqnarray}
\begin{figure}[h]
\begin{minipage}[b]{1\textwidth}
\subfigure[\label{fig1a} ]{ \includegraphics[width=.37\textwidth]%
{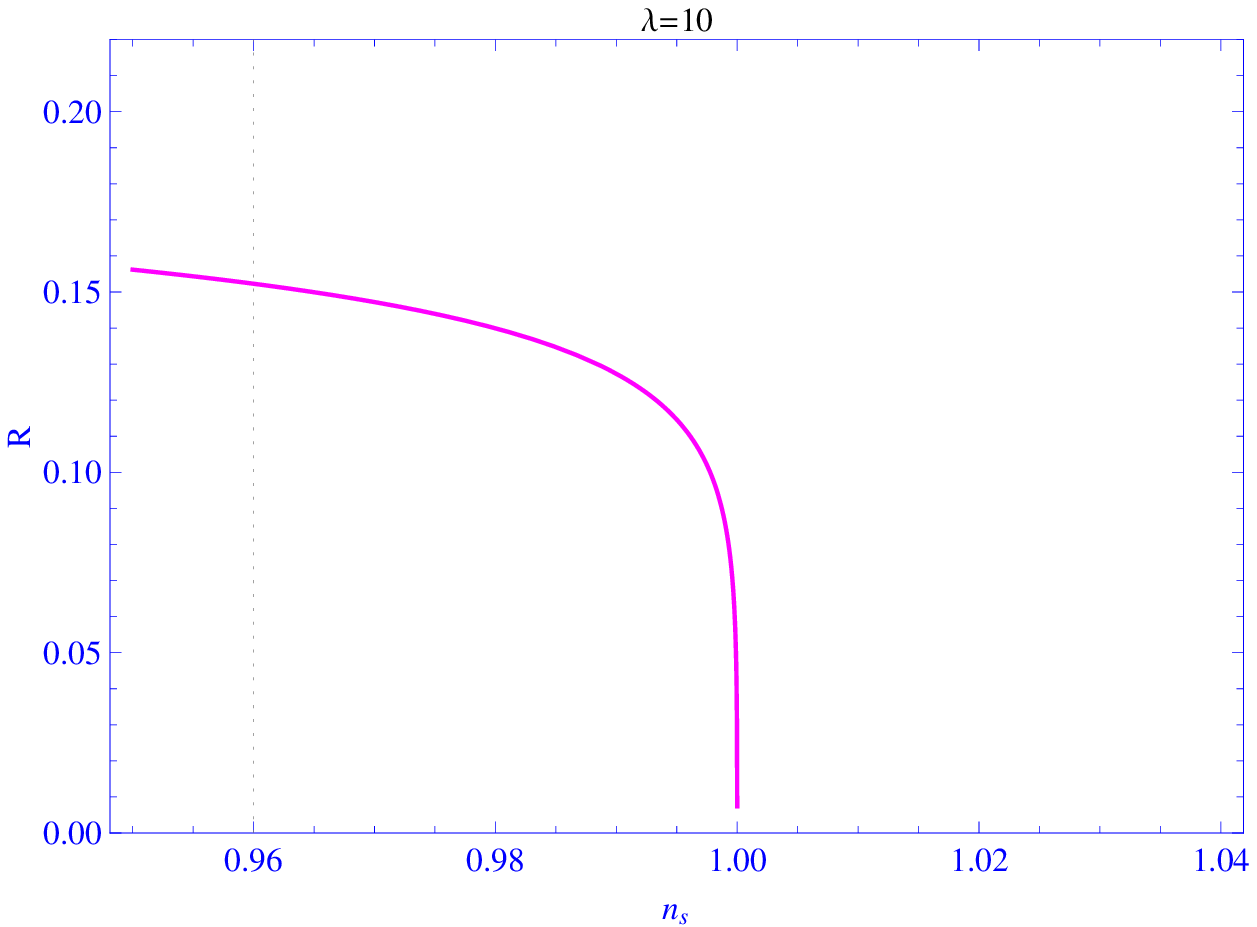}} \hspace{.2cm}
\subfigure[\label{fig1b} ]{ \includegraphics[width=.37\textwidth]%
{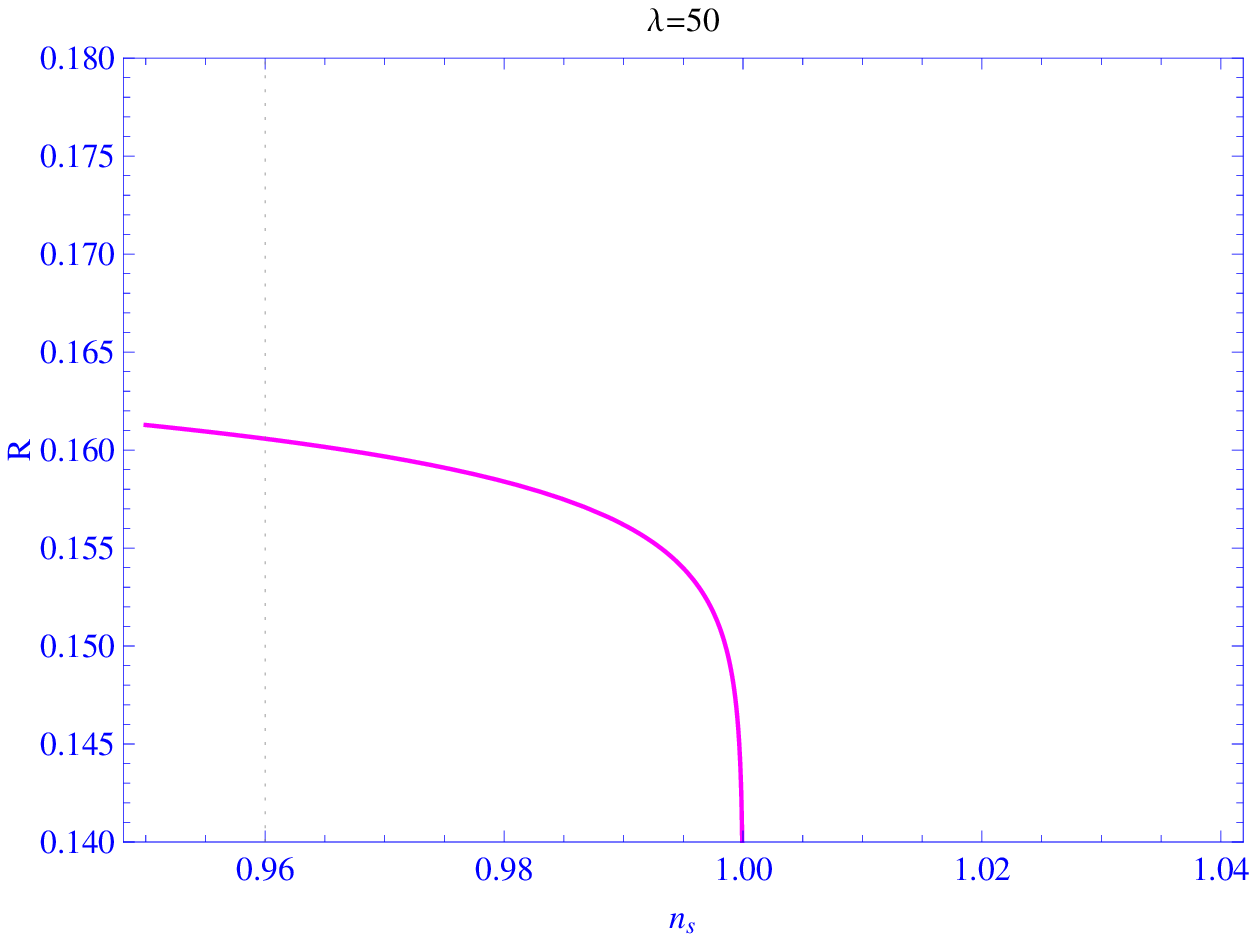}}
\end{minipage}
\caption{Scalar-tensor ratio in term of spectral index $n_s$: (a) for $\lambda=10$ and (b) for $\lambda=50$ ($A=1, \Gamma_1\propto C^{\frac{1}{6}})$).}
\end{figure}
In Fig.(8), we plot the tensor-scalar ratio  versus the spectral index   for $\lambda=10$, $\lambda=50$. We find the model is compatible with WMAP data \cite{2-i,2-m}.
\section{Conclusion}
Vector inflation model in the context of phenomenological warm inflation scenario have been studied. In this scenario, after inflation period the universe goes to radiation dominant regime and reheating epoch is avoided.  We have found the general conditions for inflation era ($\rho>\frac{2}{3}\rho_{\gamma}$)and end of this epoch ($\rho=\frac{2}{3}\rho_{\gamma}$). We have obtained the perturbation parameters of this model. These parameters are important in the observational viewpoint. Intermediate and logamediate which are exact cosmological solutions, have been studied. In this two cases we have derived slow-roll parameters and perturbation parameters in term of inflaton. We have seen that the model is compatible with WMAP7 and Planck data.
\section{Appendix}
In usual warm inflation models driven by scalar fields, there is a QFT method for   extracting the dissipation coefficient $\Gamma,$  where the inflaton is coupled to another intermediate (catalyst) field \cite{4nn}. In this method, the inflaton interacts with the catalyst heavy field which could decay into massless field which is called radiation. Following Ref.\cite{4nn}, we would like to use the same method for our model. Dissipation coefficient $\Gamma$ will be defined using this QFT method. The dynamic of our interacting system (which is denoted by field $A^{\mu}$) is in equilibrium state during inflation period. Background field $A_{\mu}$ interacts with other fields. We could define an interacting potential as
\begin{eqnarray}\label{}
V(A^{\mu}A_{\mu},X)=V(A^{\mu}A_{\mu})+V_{int}(A^{\mu}A_{\mu},X)
\end{eqnarray}
Field theory of warm vector inflation model may be described by action (\ref{1}), where the potential of this (super-cool vector inflation) model is replaced by the above potential. Other fields which are coupled to the background fields are represented by $X$. $V_{int}$ denotes the interaction of background field with other fields. We could divide this segment of potential as:
\begin{eqnarray}\label{}
V_{int}(A^{\mu}A_{\mu},X)=f(A^{\mu}A_{\mu})g(X)
\end{eqnarray}
The equation of motion for the system of triplet vector field in first order has the following form
\begin{eqnarray}\label{v1}
\partial_{\mu}\frac{\partial L_{eff}}{\partial(\partial_{\mu}B)}-\frac{\partial L_{eff}}{\partial B}-i\frac{\partial f}{\partial B}\int d^4x'\theta(t-t')[f(B(x))-f(B(x'))]<[g(X(x)),g(X(x'))]>=0
\end{eqnarray}
$L_{eff}$ is effective Lagrangian density in term of $B$. We suppose adiabatic approximation, where the inflaton $B$ is slowly varying on the timescale $\tau$
\begin{eqnarray}\label{}
\frac{\dot{B}}{B}\ll\tau^{-1}
\end{eqnarray}
where $\tau$ is response timescale for self-energy interaction \cite{4nn}. By using Taylor expansion we have
\begin{eqnarray}\label{v2}
f(B(t'))-f(B(t))=(t-t')\dot{B}\frac{\partial f}{\partial B}+...
\end{eqnarray}
The equation of motion for homogeneous field $B$ is given by
\begin{eqnarray}\label{}
\ddot{B}+(3H+\frac{\Gamma}{3})\dot{B}+V'(B_j B^j)B=0
\end{eqnarray}
From above equation and Eqs.(\ref{v1}) and (\ref{v2}) the dissipation coefficient $\Gamma$ could be defined as:
\begin{eqnarray}\label{}
\Gamma=\int d^4x'\Sigma_R(x,x')(t-t')
\end{eqnarray}
where
\begin{eqnarray}\label{}
\Sigma_R(x,x')=-i[\frac{\partial f}{\partial B}]^2\theta(t-t')<[g(X(x)),g(X(x'))]>
\end{eqnarray}
If we use a simple bi-quadratic interaction form for our model \cite{4nn}
\begin{eqnarray}\label{v3}
V_{int}(B,X)=\frac{g_1}{2}B^2X^2,
\end{eqnarray}
 The intermediate heavy fields $X$ are turned coupled to bosonic fields $L$ which  may be called radiation fields. The dissipation coefficients for some interaction structures have been derived in Refs.\cite{4nn,arj} (for warm scalar field inflation model). After the first step of interactions (\ref{v3}) (between inflaton $B$ and catalyst fields $X$), the intermediate fields $X$ decay to fields $L$ as
\begin{eqnarray}\label{v4}
V_{2int}\sim XL^2
\end{eqnarray}
where $L$ denotes the  massless (light) bosonic fields. In analogy to the calculations in Refs\cite{4nn,arj}, if the two-stage interaction structures are as Eqs.(\ref{v3}),(\ref{v4}), one can expect a form such as
\begin{eqnarray}\label{}
\Gamma=\Gamma_0\frac{T^3}{B^2}
\end{eqnarray}
for dissipation coefficient in low temperature limit.
The above form of dissipation coefficient is presented as a possible type of dissipative term
but there could be other forms. We have used this form of dissipation in sections (III) and (IV).

\end{document}